\documentstyle[12pt]{article}

\begin{document} 

\begin{titlepage}

\hrule 
\leftline{}
\leftline{Chiba Univ. Preprint
          \hfill   \hbox{\bf CHIBA-EP-109}}
\leftline{\hfill   \hbox{hep-th/9810167}}
\leftline{\hfill   \hbox{October 1998}}
\vskip 5pt
\hrule 
\vskip 1.0cm

\centerline{\large\bf 
Quark Confinement and Deconfinement in QCD 
} 
\vskip 0.5cm
\centerline{\large\bf  
from the Viewpoint of
}
\vskip 0.5cm
\centerline{\large\bf  
Abelian-Projected Effective Gauge Theory
}

\vskip 1cm

\centerline{{\bf 
Kei-Ichi Kondo$^{1,}{}^{\dagger}$
}}  
\vskip 1cm
\begin{description}
\item[]{\it \centerline{ 
$^1$ Department of Physics, Faculty of Science, 
Chiba University,  Chiba 263-8522, Japan}
  }
\item[]{\centerline{$^\dagger$ 
  E-mail:  kondo@cuphd.nd.chiba-u.ac.jp }
  }
\end{description}

\centerline{{\bf Abstract}} 
We give another derivation of quark confinement in QCD from the
viewpoint of the low-energy effective Abelian gauge theory of QCD
obtained via Abelian projection.   It is based on the recently
discovered Berezinskii-Kosterlitz-Thouless phase transition in the
four-dimensional Abelian gauge theory.  Moreover, we show that
there exists a critical gauge coupling constant in QCD, above which
confinement set in and below which there is no confinement. In a
SU(N) gauge theory with $N_f$ flavored fermions, we argue that this
leads to a critical value of fermion flavors $N_f^c$ for the
confinement as well as the chiral symmetry breaking, which separates
the deconfining and chiral symmetric phase from the confining and
chiral symmetry breaking phase.  A finite-temperature deconfinement
transition is also discussed briefly.


\vskip 0.2cm
\hrule  

\vskip 0.5cm  

$^*$ To be published in Phys. Lett. B.

\end{titlepage}

\baselineskip 23pt

\newpage
\par

Proving quark confinement in  quantum chromodynamics
(QCD) is a long-standing problem in theoretical particle physics.
In a series of papers \cite{KondoI,KondoII,KondoIII,KondoIV}, the
author claimed that a novel {\it reformulation} of Yang-Mills (YM)
theory leads to quark confinement in  QCD.  The reformulation is
based on the identification of the YM theory as a perturbative
deformation of a novel topological quantum field theory (TQFT).  The
TQFT is suggested from the gauge-fixing and the Faddeev-Popov ghost
terms in the maximal Abelian gauge (MAG)
\cite{KLSW87}.  The MAG is a choice of partial gauge fixing which
realizes in the context of quantized gauge field theory an idea of
{\it Abelian projection} proposed by 't Hooft
\cite{tHooft81}.   After the Abelian projection, the original
non-Abelian gauge theory is rewritten into the Abelian gauge theory
with magnetic monopoles and adjoint matter fields.  
The TQFT in question is a field theory which describes the
topologically non-trivial (background) gauge field configurations
leading to non-vanishing magnetic monopole current loop (due to
topological conservation law) in four space-time dimensions. As
demonstrated in
\cite{KondoII}, such topological field configurations in the
functional integral give the dominant contribution to the string
tension $\sigma$ (or area decay of the Wilson loop), while the
remaining part is identified with the perturbative contribution
around such topological configurations (leading to at most perimeter
decay).  This result supports the dual superconductor picture of QCD
vacuum
\cite{Nambu74}.
\par
The term `Abelian projection' is somewhat misleading,
since it sounds as if off-diagonal parts are neglected a priori. 
Indeed, such a standpoint was taken in the early stage of
investigation on quark confinement \cite{EI82,Suzuki88}.  It is a
trivial statement that the Abelian gauge theory is obtained
 from an non-Abelian gauge theory by neglecting the off-diagonal
gauge fields. Note that the Abelian projection
here takes fully into account the off-diagonal parts by integration
(in the sense of Wilsonian renormalization group), which reduces to
the {\it renormalization} of the effective Abelian gauge theory as
shown in 
\cite{KondoI}.
\par
We adopt  as a criterion of quark confinement  
the area law of the {\it large} (non self-intersecting)
Wilson loop (large compared with the typical size of QCD).  
What we have shown in the previous papers and what we are going to
show in this paper are concerned with the physics in the low-energy
or long-distance region of QCD.  Therefore, we don't say anything
novel about the high-energy or short-distance region of QCD where the
asymptotic freedom is essential.  This way of thought constitutes
the core of Abelian dominance in low-energy physics of
QCD \cite{EI82}.  Actually, it turns out that Abelian
dominance is lost in the high-energy or short-distance region of
QCD. 
\par
The above reformulation does not necessarily guarantee the
non-existence of additional non-perturbative effects (coming from the
remaining gluon self-interactions) which might invalidate 
the proof claimed above, although the validity of the identification
is checked by numerical simulations.
In this paper, in order to get rid of this doubt, we give
another derivation of quark confinement by combining previous
works
\cite{KondoI} and \cite{KondoII,KondoIII,KondoIV}. 
\footnote{
The main idea is still the dual superconductor picture of QCD
vacuum where the magnetic monopole plays the crucial role. 
However, this does not imply that the instantons
and other topological non-trivial configurations such as vortex have
nothing to do with quark confinement.  
Exact separation of monopole from other configuration is
meaningless, since they are overlapped with each other.
}


First, we consider a non-Abelian gauge theory described by the
Yang-Mills (YM) theory with a compact (semi-simple) non-Abelian gauge
group
$G$.  Then we decompose the gauge field $\cal A_\mu$ into the
diagonal
$a_\mu$ and off-diagonal $A_\mu$ pieces (Cartan decomposition).  The diagonal gauge field belongs to the
maximal torus group $H$ of
$G$, while the off-diagonal gauge field to the coset $G/H$. 
The Abelian projection is realized as a partial gauge fixing which
fixes the gauge degrees of freedom corresponding to  $G/H$.
Under the residual gauge transformation of $H$, the
diagonal gluon fields transform as Abelian gauge fields, whereas
the off-diagonal gluons transform as adjoint matter fields.
Second, we obtain the Abelian gauge theory with a gauge group
$H$ by integrating out all the off-diagonal gauge fields belonging to
$G/H$ \cite{QR97}.  This is possible by introducing the auxiliary
antisymmetric tensor field $B_{\mu\nu}$. 
  We called the Abelian gauge theory obtained in this way the
Abelian-projected effective gauge theory (APEGT)
\cite{KondoI}.   
\par
A quite important observation is that the APEGT  obtained  in this
way is considered as the low-energy effective gauge theory (LEEGT) of
the original non-Abelian gauge theory, e.g. QCD.  This is because the
off-diagonal gluons become massive after the MAG, as shown by direct
numerical simulations
\cite{ASu97} and an analytical consideration \cite{KondoII}. 
Therefore, if the integration over the coset components
performed above is identified with the integration over the
high-energy mode in the sense of Wilsonian renormalization group,
the resulting Abelian gauge theory is regarded as the LEEGT which is
meaningful in the low energy-momentum region ($p < O(m_A)$) or in
the length scale $\ell$ larger than the inverse of the off-diagonal
gluon mass
$m_A$ ($\ell > O(m_A^{-1})$).
This is another important point which is sometimes misunderstood by
the literatures.
\par
The  APEGT of the YM theory with a gauge group $G$ obtained
according to the above procedures 
has the action \cite{KondoI} (in 3+1 space-time dimensions),
\footnote{
Although, in the paper \cite{KondoI}, the actual calculation has been
presented only for $G=SU(2)$, the extension to $G=SU(N)$ is
straightforward, at least in the one-loop level, as demonstrated in
\cite{KS98}. }
\begin{eqnarray}
 S_{APEGT} = \int d^4x \left[ - {1+z_f \over 4g^2}
f_{\mu\nu}f^{\mu\nu}
 - {1+z_B \over 4}g^2 B_{\mu\nu}B^{\mu\nu} 
 + {1 \over 4} z_k B_{\mu\nu}
\epsilon_{\mu\nu\rho\sigma}f^{\rho\sigma} 
 + \cdots \right] ,
 \label{APEGT}
\end{eqnarray}
where $f_{\mu\nu}$ is the Abelian field strength
$f_{\mu\nu}:=\partial_\mu a_\nu - \partial_\nu a_\mu$, 
$B_{\mu\nu}$ is the rank 2 antisymmetric tensor field and
the dots denote  ghost self-interaction terms and
 higher derivative terms \cite{KondoI}.
The renormalization factors are given by
\begin{eqnarray}
  z_f &=&  - {22 \over 3} C_2(G) {g^2 \over 16\pi^2} \ln \mu ,
  \label{zf}
  \\
  z_B &=& + 2 C_2(G) {g^2 \over 16\pi^2} \ln \mu ,
  \\
  z_k &=& + 4 C_2(G) {g^2 \over 16\pi^2} \ln \mu ,
\label{z}
\end{eqnarray}
where $\mu$ is the relevant scale of the theory and $C_2(G)$ is
the quadratic Casimir operator,
$
 C_2(G) \delta^{AB} = f^{ACD}f^{BCD}.
$
The APEGT (\ref{APEGT}) is an Abelian gauge theory with the gauge
group
$H_e \times H_m$ where $H_e=U(1)^{N-1}$ denotes the maximal torus
subgroup of  $G$ (the subscript e is an abbreviation
of electric) and
$H_m$ is the same group as
$H_e$ but identified with the gauge group for the dual gauge field
$b_\mu$ (the subscript m is an abbreviation of magnetic).
Here the dual gauge field $b_\mu$ is obtained from the Hodge
decomposition of  $B_{\mu\nu}$ in 3+1
dimensions,
$
  B_{\mu\nu} = b_{\mu\nu} + \tilde \chi_{\mu\nu} , 
$
$
 b_{\mu\nu} := \partial_\mu b_\nu - \partial_\nu b_\mu ,
$
$
  \tilde \chi_{\mu\nu}= {1 \over 2}
\epsilon_{\mu\nu\alpha\beta}
 (\partial^\alpha \chi^\beta - \partial^\alpha \chi^\beta) .
$
Consequently, the dual gauge field gets coupled with the
monopole current $k_\mu$ as
\begin{eqnarray}
 \int d^4x \  B_{\mu\nu} \tilde f^{\mu\nu} 
 \equiv \int d^4x \  \tilde B_{\mu\nu} f^{\mu\nu} 
 = \int d^4x  \  b_\mu  k^\mu + ... ,
 \label{monoc}
\end{eqnarray}
where the non-vanishing magnetic monopole current is deduced as
violation of Bianchi identity for the Abelian part,
$
  k^\mu := \partial^\nu \tilde f_{\mu\nu},
$
where
$
 \tilde f_{\mu\nu} :=  {1 \over 2} 
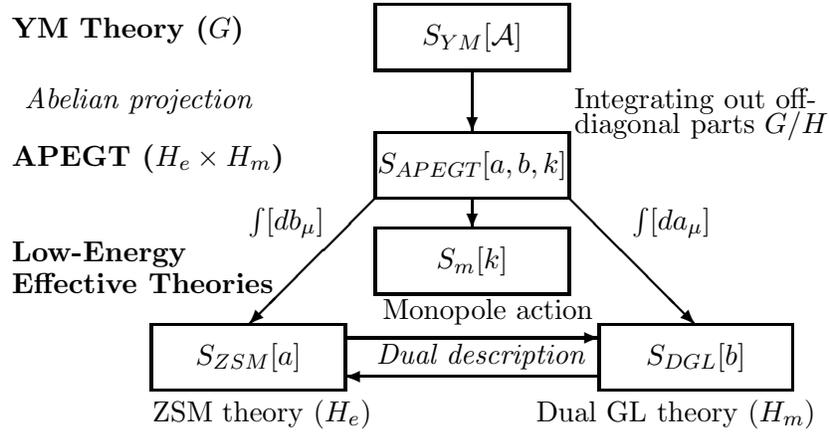
\epsilon_{\mu\nu\rho\sigma}f^{\rho\sigma} .
$

\par
\vskip 0.5cm
\begin{center}
\unitlength=0.85cm
\thicklines
\small
 \begin{picture}(12,6)
 \put(5.5,5){\framebox(3,1){$S_{YM}[{\cal A}]$}}
 \put(7,5){\vector(0,-1){1}}
 \put(-0.2,5.5){\bf YM Theory ($G$)}
 \put(8.6,4.4){Integrating out off-}
 \put(8.6,4.0){diagonal parts $G/H$}
 \put(0.0,4.4){\it Abelian projection}
 \put(5.5,3){\framebox(3,1){$S_{APEGT}[a,b,k]$}}
 \put(7,3){\vector(0,-1){0.5}}
 \put(-0.2,3.5){\bf APEGT ($H_e\times H_m$)}
 \put(5.5,1.5){\framebox(3,1){$S_{m}[k]$}}
 \put(5.6,1.1){Monopole action}
 \put(5.5,3){\vector(-1,-1){2}}
 \put(3.5,2.5){$\int[db_\mu]$}
 \put(-0.2,2){\bf Low-Energy}
 \put(-0.2,1.5){\bf Effective Theories}
 \put(2,0){\framebox(3,1){$S_{ZSM}[a]$}}
 \put(8.5,3){\vector(1,-1){2}}
 \put(9.5,2.5){$\int[da_\mu]$}
 \put(9,0){\framebox(3,1){$S_{DGL}[b]$}}
 \put(5,0.8){\vector(1,0){4}}
 \put(9,0.2){\vector(-1,0){4}}
 \put(5.5,0.4){\it Dual description}
 \put(2.0,-0.5){ZSM theory ($H_e$)}
 \put(8,-0.5){Dual GL theory ($H_m$)}
 \end{picture}
\end{center}

\par
The APEGT  can give  dual descriptions of the low-energy physics in
the following sense. After eliminating $a_\mu$ field by integration,
the APEGT yields the dual Ginzburg-Landau theory (DGL)
with an action $S_{DGL}[b]$ which is written in terms of the dual
gauge field $b_\mu$ alone.  On the other hand, the APEGT is rewritten
into another Abelian gauge theory with an action
$S_{ZS}[a]$ written in terms of the $a_\mu$ field alone after
integrating out the $b_\mu$ field. 
\footnote{
The theory was called the Zwanziger--Suzuki-Maedan (ZSM) theory,
since it was obtained by Suzuki for $G=SU(2)$ and Maedan-Suzuki  for
$G=SU(3)$
\cite{Suzuki88} using the Zwanziger formalism
\cite{Zwanziger71}. 
However, the derivation of ZS theory was based on the ad hoc
treatment of the off-diagonal parts  under the assumption of
Abelian dominance
\cite{EI82} and assumption of the monopole
condensation which leads to the mass for the dual gauge field as a
result of spontaneous breakdown of the dual gauge symmetry $H_m$
(dual Meissner effect or dual Higgs mechanism). 
These points are improved in \cite{KondoI}.
}
In this sense APEGT  is an interpolating theory between two theories
just described (We can also write the monopole current action
$S_m[k]$).  The duality of the APEGT is explicit in the action
(\ref{APEGT}), since at the level of field equation,
$
 B^{\mu\nu} =  \tilde f^{\mu\nu},
$
or
$
 f^{\mu\nu} =  \tilde B^{\mu\nu} ,
$
up to a multiplicative constant (renormalization factor).

\par

It is possible to discuss the confinement through the DGL theory
which is derived from APEGT as a  LEEGT of QCD.  However,
in order to conclude quark confinement in this
scenario (dual superconductivity \cite{Nambu74}), it is necessary to
show that monopole condensation really takes place.  The analytical
studies on quark confinement so far have been based on the
assumption of monopole condensation in QCD.   Although it is
actually shown to occur at least in the strong coupling region based
on the energy (action)-entropy argument in the lattice regularization
\cite{KondoI}, we will adopt another path in this paper.  
In (\ref{APEGT}), $g$ denotes the bare
coupling constant and hence the prefactors
$z_f, z_B$ imply the renormalization effect.  
The calculated prefactor $z_B$ shows that the DGL theory
$S_{DGL}[b]$ does not exhibit asymptotic freedom, as expected.

\par
On the other hand, eliminating the field $B_{\mu\nu}$ from the
interpolating theory (\ref{APEGT}) by a Gaussion integration, we
obtain up to one-loop level
\footnote{
Note that this action is valid just before the monopole condensation
occurs.  After the monopole condensation, the action should be
modified. We consider approaching the confinement phase from the
deconfinement phase.
}
\begin{eqnarray}
 S_{} = - \int d^4x {1 \over 4g(\mu)^2} f_{\mu\nu}f^{\mu\nu}
 + \cdots ,
 \label{APEGT2}
\end{eqnarray}
which is nothing but the Abelian gauge theory with a gauge group
$H=U(1)^{N-1}$.  However, a remarkable difference from the true
Abelian gauge theory is that the coupling constant
$g(\mu)$ runs according to the same renormalization group beta
function as the original YM theory \cite{KondoI} (see
eq.~(\ref{zf})),
\begin{eqnarray}
  \beta(g) := \mu {dg(\mu) \over d\mu} 
= - {b_0 \over 16\pi^2} g(\mu)^3, 
\quad  b_0 = {11C_2(G) \over 3} > 0 .
\label{beta}
\end{eqnarray}
In this sense, this LEEGT reflects
correctly the asymptotic freedom of the original YM theory.
Indeed, the beta function (\ref{beta}) leads to the running coupling,
\begin{eqnarray}
  g(\mu)^2 =  {g(\mu_0)^2 \over 1
  + {b_0 \over 8\pi^2} g(\mu_0)^2 \ln {\mu \over \mu_0}} .
  \label{rc}
\end{eqnarray}
As the length scale is larger and larger, the running gauge coupling
becomes larger and eventually goes to infinity in the one-loop
level, whereas, at two-loop level, it converges to the infrared fixed
point $\alpha=\alpha_*$ \cite{ATW96} for $\alpha:={g^2 \over 4\pi}$.

\par
The purpose of this paper is
to derive quark confinement in QCD based on the effective
Abelian gauge theory (\ref{APEGT2}) derived from the APEGT without
assuming monopole condensation.   
We avoid to use the Zwanziger formalism which introduces an
arbitrary unit four vector
$n_\mu$. Hence we do not need to worry about the ambiguity how to
choose the vector $n_\mu$.
The same effect as monopole
condensation is supplied with the vortex condensation based on the
reformulation of the Abelian gauge theory (\ref{APEGT2}) as a
 deformation of another TQFT \cite{KondoIII}.  
This approach can greatly simplifies the proof (e.g., we can avoid
the tedious instanton calculations) and avoid the ambiguity
just mentioned.  In this strategy, we are dealing with the
Abelian gauge theory alone.  Hence the doubt on the
identification of the perturbative sector does not occur,
since the Maxwell theory has no self-interaction among the gauge
fields. This is another advantage of this approach.  Moreover, this
view gives an insight  into other interesting problems, e.g. quark
deconfinement transition and chiral symmetry breaking (CSB)
discussed below.   Technical details will be given elsewhere
\cite{Kondo99}.

\par
Now we show that the quark confinement is realized in the
non-Abelian gauge theory in the sense that the expectation value of
the  Wilson loop obeys the area law for sufficiently large
(non self-intersecting) Wilson loop.  In other words, the dominant
contribution of the static quark potential between a quark and an
anti-quark is given by the linear potential in the large separation
(In addition, there is the Coulomb potential  and a
constant part which expresses the self-energy contribution).   
\footnote{
It should be
remarked that we do not conclude anything new about the short
distance behavior of the Wilson loop or static potential, because the
following argument is based on the APEGT which is meaningful only in
the large distance or low-energy region as explained above.  
}
The information on the high-energy physics is incorporated
into the APEGT through the running coupling constant (\ref{rc}). 
This is quite important to prove quark confinement based on the
effective Abelian gauge theory.  
\par
In fact, if we regard the above theory (\ref{APEGT2}) with the
LEEGT of the YM theory, it turns out that the quark confinement in
QCD follows immediately from the result of the previous paper
\cite{KondoIII}, i.e. the existence of a confining phase in QED. 
In  \cite{KondoIII} it was shown that the
four-dimensional Abelian gauge theory has a confining phase in the
strong coupling region
$g>g_{c}$ where the fractional charge is confined by the linear
static potential.  The steps are as follows.
The theory (\ref{APEGT2})  has the residual Abelian gauge invariance
$H$.  In order to quantize this theory, we must fix also the
residual Abelian gauge degree of freedom.  We adopt a
manifestly covariant gauge fixing of Lorentz type.  For a choice of
gauge-fixing parameter
$\beta=-2$ \cite{KondoIII}, the dimensional reduction occurs.  Then
the critical coupling $g_c$ is determined from the
Berezinskii-Kosterlitz-Thouless (BKT) transition temperature
$T_{BKT}=8\pi$ of the two-dimensional O(2) nonlinear sigma model
(NLSM):
$g_{c}=\pi$. The weak coupling region $g<g_c$ is the conformal
(Coulomb) phase.
The above steps are shown schematically as follows.
\par
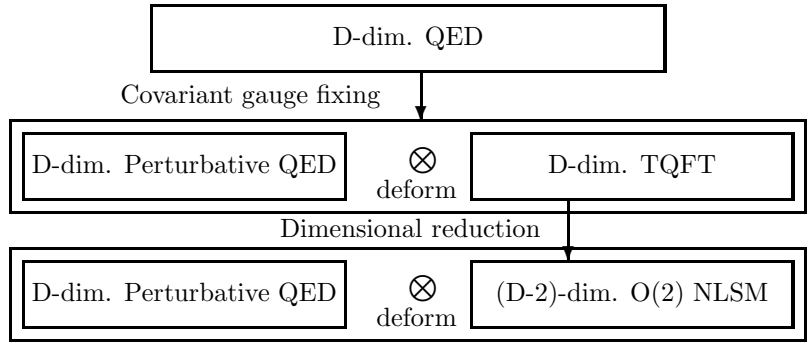
\begin{center}
\unitlength=0.85cm
\thicklines
\footnotesize
 \begin{picture}(12,6)
 \put(2,5){\framebox(8,1){D-dim. QED}}
 \put(6.2,5){\vector(0,-1){0.8}}
 \put(1.5,4.5){Covariant gauge fixing}
 \put(-0.2,2.8){\framebox(12.4,1.4){}}
 \put(0,3){\framebox(5,1){D-dim. Perturbative QED}}
 \put(6,3.5){$\bigotimes$}
 \put(5.5,3){deform}
 \put(7,3){\framebox(5,1){D-dim. TQFT}}
 \put(8.5,3){\vector(0,-1){1}}
 \put(4.0,2.4){Dimensional reduction}
 \put(-0.2,0.8){\framebox(12.4,1.4){}}
 \put(0,1){\framebox(5,1){D-dim. Perturbative QED}}
 \put(6,1.5){$\bigotimes$}
 \put(5.5,1){deform}
 \put(7,1){\framebox(5,1){(D-2)-dim. O(2) NLSM}}
 \end{picture}
\end{center}

Furthermore, the evaluation of the Wilson loop is performed using
the  equivalence of O(2) NLSM with (neutral) Coulomb gas, sine-Gordon
model and the massive Thirring model in two dimensions. 
This fact will shed light on the confining string picture
\cite{Polyakov97} in the low-energy region of QCD.

\par
Note that the Abelian gauge theory (\ref{APEGT2}) is different from
the usual perturbative QED in the following points.
First, the Abelian group $H$ is compact,
since it was embedded in the compact (non-Abelian) group $G$, while
the U(1) gauge group in the perturbative QED is considered to be
non-compact. Due to compactness or periodicity, the Abelian
gauge field can have the topologically non-trivial configuration
(vortices). 
It is the periodicity why the fractional charge is confined while
the integral charge is not confined.
Second, the {\it true} Abelian gauge theory can not be asymptotic
free.  That is to say, in the infrared limit $p \downarrow 0$ or long
distance limit $\ell \uparrow \infty$,  the effective charge
$\alpha(p)$ goes to zero.  Therefore, QED
inevitably goes into the Coulomb phase. On the contrary, the running
gauge coupling (\ref{rc}) of (\ref{APEGT2}) increases monotonically
in the length scale $\ell$ and eventually becomes larger than the
critical coupling
$\alpha_c$ with increasing $\ell$.   Thus, in the low-energy
region
$\mu<\mu_c$ (or $\ell > \ell_c$) compatible with $\mu<O(m_A)$ (or
$\ell > O(m_A^{-1})$), the theory (\ref{APEGT2}) lies in the
confining phase. The two critical scales $m_A, \mu_c$ are expected
to be the same order of magnitude ($\sim$ 1GeV). 
Thus the critical scale $\mu_c$ where the confinement sets in is
given by 
\begin{equation}
\alpha(\mu_c)=\alpha_c=\pi/4 .
\end{equation}

\par
The above consideration should be compared with the previous work
\cite{KondoII} where quark confinement has been derived by directly
studying the non-Abelian gauge theory without considering the LEEGT.
The above consideration supplements the previous treatment
\cite{KondoII}. 
By the existence of the scale $\mu_c$, the separation of the gauge
variable $\cal A_\mu$ into $\Omega_\mu$ and $\cal V_\mu$ performed in
\cite{KondoII} has the following meaning. The background piece
$\Omega_\mu$ represents  the low-energy (or long-distance) mode
$\mu<\mu_c$ while the perturbative piece $\cal V_\mu$  the
high-energy (or short distance) mode
$\mu>\mu_c$ in  QCD.
\par
 {}From the result of \cite{KondoIII} on the Abelian Wilson loop,  
therefore, the above consideration gives another proof of quark
confinement in the sense of area law of the diagonal Wilson loop
claimed in \cite{KondoII}. 
Finally, the area law of the non-Abelian Wilson
loop is shown as follows.  The non-Abelian
Wilson loop is rewritten as a path integral of the
Abelian Wilson loop via a non-Abelian Stokes theorem as exemplified
in the paper
\cite{KondoIV}.  Therefore, the area law of the non-Abelian Wilson
loop follows from that of the Abelian-projected Wilson loop which
is written in terms of the Abelian (or diagonal) part alone, see
\cite{Kondo99}.

\par

In what follows, we consider a few applications of the above result.
First, we consider a SU(N) gauge theory with $N_f$ flavored
fermions. If the massless
fermions are present in the above consideration, the Wilson loop
loses its meaning as a criterion of quark confinement.  This is
because the virtual pairs of a fermion and an antifermion are
produced from the vacuum and they break the string of inter-quark
flux leading to the perimeter decay of the Wilson loop.  Then the
static potential is saturated to be a constant in the asymptotic
region. However, if the fermion mass
$m_f$ is non-zero (even if small), the above criterion is still
meaningful. The reason is as follows. Once the dynamical fermion
mass $m_f$ is produced, the fermions decouple below this scale,
leaving the pure gauge theory behind. Therefore, the linear
potential 
$V(R)=\sigma R$ survives the
inclusion of light quarks for the length scale 
$m_A^{-1} < \ell< m_f \sigma^{-1}$ ($m_f^{-1}\sigma < p < m_A$). 
\footnote{
Here we have implicitly assumed that the string tension does not
change substantially despite the inclusion of dynamical fermions in
the relevant region.
}
If the theory with fermions is asymptotic free and hence
the running coupling constant is monotonically increasing toward the
infrared region, a criterion of quark confinement is given by
$
  \alpha(\mu)  > \alpha_c = {\pi \over 4}
$
as long as the fermion is massive ($m_f\not=0$).

On the other hand, it is known that the chiral symmetry is
spontaneously broken above the critical coupling
$\alpha>\alpha_c'$. For example, the  SU(N) gauge theory has the
critical coupling for the chiral symmetry breaking (CSB),
$
 \alpha_c' = {\pi \over 3C_2(R)} = {2N \over 3(N^2-1)}\pi ,
$
according to the naive Schwinger-Dyson (SD) equation \cite{ATW96}.
This implies that the chiral symmetry is broken for not too
large $N_f$ below a critical value $N_f^c$,
$N_f<N_f^c$. 
This leads to a critical value of fermion flavors
$N_f^c$ for the confinement as well as the  CSB, which separates
the deconfinement and chiral symmetric phase as follows.   
For $N=2,3$, the critical coupling $\alpha_c'$ of CSB is
greater than or equal to the critical coupling
$\alpha_c$ of  quark confinement,
$\alpha_c' \ge \alpha_c$ (In fact, for $N=2$, $\alpha_c'=4\pi/9$ and
for $N=3$
$\alpha_c'=\pi/4$ which happens to coincide with the critical
coupling $\alpha_c$). 
\footnote{
In the region $\alpha_c'<\alpha<\alpha_c$ (which is expected to
occur for large $N$), the fermion is massive and the chiral symmetry
is broken, but quark is not confined. 
}
In the region 
$\alpha_c< \alpha_* <\alpha_c'$ ($\alpha_*$: the infrared fixed
point), although the fermion remains massless in the naive treatment
without topological contribution, the vortex plasma (monopole
condensation) can induce the CSB which causes quark confinement. It
is also known that monopole condensation enhances the CSB
\cite{SST95}. If this observation is correct, the CSB and
confinement can occur simultaneously.

We can estimate the critical number of flavors $N_f^c$ for the
deconfinement/confinement transition as follows. For SU(N) QCD with
$N_f$ fermions in the fundamental representation,  the
renormalization group beta function is given by
\begin{eqnarray}
  \beta(\alpha) := \mu {\partial \alpha \over \partial \mu}
  = - b \alpha^2(\mu) - c
\alpha^3(\mu)
\cdots ,
\end{eqnarray}
with the first
two coefficients,
\begin{eqnarray}
  b = {1 \over 6\pi}(11 N - 2N_f), \quad
  c = {1 \over 24\pi^2} \left( 34 N^2 - 10 N N_f - 3 {N^2-1 \over N}
N_f \right) .
\end{eqnarray}  
Note that, above the value $N_f=(11/2)N$ ($b<0,c<0$)
($N_f=11$ for $N=2$ and $N_f=16{1 \over 2}$ for $N=3$), the theory
is not asymptotic free, i.e., the beta function is positive
for any
$\alpha$,
$\beta(\alpha)>0$.  Below this value ($b>0,c<0$), the infrared fixed
point
$\alpha_*$ exists where the fixed point is given by $\alpha_* = -
b/c>0$.  From this relation, we have
$\alpha_*={4\pi(11N^2-2NN_f) \over (13N^2-3)N_f-34N^3}$,
so that the above relation is meaningful only when
$N_f>34N^3/(13N^2-3)$,
since $\alpha_*>0$ only if this condition is satisfied (otherwise,
$\alpha_*<0$). 
 Note that $\alpha_*$ increases monotonically as the flavor $N_f$ is
decreasing and $\alpha_* \uparrow +\infty$ as
$N_f \downarrow 34N^3/(13N^2-3)$. 
For large $N$, $N_f$ approaches $(34/13)N$.  
For $N_f<34N^3/(13N^2-3)$, $\beta(\alpha)<0$ and the theory is
asymptotic free (neglecting higher orders $O(\alpha^5)$). This
argument gives the lower bounds,
$N_f^c \ge 5.55$ for $N=2$ and $N_f^c \ge 8.05$ for $N=3$.
The relationship between the critical flavor and the critical
coupling for the deconfinement/confinement transition is obtained
by putting $\alpha_*=\alpha_c$ as 
\begin{eqnarray}
  N_f^c = {2(22\pi + 17\alpha_c N)N^2 \over 8\pi N + 
  \alpha_c (13N^2-3)} .
\end{eqnarray}
The critical coupling $\alpha_c=\pi/4$ leads to
$N_f^c=8.64$ for $N=2$ and $N_f^c=11.9$ for $N=3$, under the
assumption that the CSB has already occured.  Therefore, the critical
number of flavors is estimated:
$8 \ge N_f^c \ge 6$ for $N=2$ and $12 \ge N_f^c \ge 8$ for $N=3$. 
These results are consistent with the results of lattice
simulations \cite{IKKSY98}.
\par
Here we have used the result
of SD gap equation for the CSB \cite{ATW96}.   However, in order to
really understand the relationship between confinement and chiral
symmetry breaking, it is desirable   to derive both of them on
equal footing from QCD.   This  will be given in a forthcoming paper
\cite{Kondo99}.

\par
Next, we consider a finite-temperature deconfinement transition
briefly.  The effective coupling at finite temperature behaves as
$
 \alpha_{EFF}(T) \sim 1/(b_0 \log (T/\Lambda)) .
$
If the temperature $T$ is sufficiently high so that $\alpha_{EFF}(T)
<\alpha_c$, quark confinement can not occur based on the above
arguments.  Due to  monotonicity of $\alpha_{EFF}(T) $, therefore,
the confinement--deconfinement transition point is determined by the
equation, 
$
 \alpha_{EFF}(T) = \alpha_c = \pi/4.
$

\par
The above strategy for proving quark confinement can be applied to
the non-Abelian gauge theory defined in the $D=d+1$ dimensional
space time for arbitrary
$D>2$.
For example, if we apply the above method to the 2+1 dimensional
non-Abelian gauge theory, the 2+1 dimensional TQFT is ,
after the dimensional reduction, reduced to the 0+1 dimensional NLSM
which   is  not a model of the quantum field theory, but that of
quantum mechanics.  It is nothing but the plane rotator model in
quantum mechanics. In quantum mechanics, the phase transition does
not occur and the theory lives in one phase.  The phase
should correspond to the confining phase of the 2+1 dimensional
non-Abelian gauge theory.  However, the Wilson loop can not be
defined in 0+1 dimensional spacetime.  We will need other criterion
of quark confinement.

\par
Obviously, it is important to extend our argument to the two-loop
level by deriving the APEGT of QCD at two-loop level
\cite{Kondo99,KS98}. Furthermore, it is desirable to extend the
arguments given for
$G=SU(2)$ in this paper to $G=SU(N), N>2$.
In a subsequent paper \cite{Kondo99} we will deal  with the
generalized BKT transition corresponding to $G=SU(N)$.  Moreover, the
area law of the non-Abelian Wilson loop will be derived based
on fermionization of generalized sine-Gordon model \cite{Kondo99}.
\par
In \cite{KondoII}, quark confinement has been explained as a
result of instantons in two-dimensional NLSM.  
However, this is not inconsistent with the vortex picture for
the effective Abelian gauge theory given in this paper, since it
turns out that an instanton is composed of a vortex and an
anti-vortex.  The Abelian and non-Abelian quark confinement will be
understood in a unified treatment \cite{Kondo99}.

This work is supported in part by
the Grant-in-Aid for Scientific Research from the Ministry of
Education, Science and Culture (10640249).


\end{document}